\documentclass{elsart}
\usepackage{amsfonts}
\usepackage{amsmath}
\usepackage{amssymb}
\usepackage{graphicx}
\usepackage{wick}

\journal{Nucl. Phys. A}

\newcommand\openone{\leavevmode\hbox{\small1\normalsize\kern-.33em1}}

\begin{document}

\begin{frontmatter}

\title{The effect of nonlocal confining kernels on magnetic chiral condensates}

\author[CFTP]{R. Gonz\'{a}lez Felipe}, \author[CFIF]{G. M. Marques},
\author[CFIF]{J. E. Ribeiro}

\address[CFTP]{Departamento de F\'{\i}sica and Centro de
F\'{\i}sica Te\'orica de Part\'{\i}culas, Instituto Superior T\'{e}cnico, Av. Rovisco Pais,
1049-001 Lisboa, Portugal}

\address[CFIF]{Departamento de F\'{\i}sica and Centro de F\'{\i}sica das
Interac\c{c}\~{o}es Fundamentais, Instituto Superior T\'{e}cnico, Avenida Rovisco Pais,
1049-001 Lisboa, Portugal}

\begin{abstract}
The physics of spontaneous chiral symmetry breaking in the case of the
simultaneous presence of a magnetic field and a fermionic quartic interaction
is discussed for both local and nonlocal kernels in 2+1 and 3+1 dimensions. The
approach is based on the use of Valatin-Bogoliubov canonical transformations,
which allow, in the absence of fermionic quartic terms, to completely
diagonalize the Hamiltonian and construct the vacuum state.

\vspace{5mm} \noindent \emph{PACS}: 12.38.Aw, 11.30.Rd, 11.10.Kk
\end{abstract}

\end{frontmatter}

\section{Introduction}

During the past few years, the behavior of QCD and, more specifically, the
problem of dynamical chiral symmetry breaking in the presence of a strong
magnetic field have attracted some
attention~\cite{Klimenko:1991he,Gusynin:va,Kabat:2002er}. It has been shown
that a constant magnetic field acts as a strong catalyst of dynamical chiral
symmetry breaking, thus leading to the generation of a fermion
mass~\cite{Klimenko:1991he,Gusynin:va}. The physics behind this effect is easy
to understand: the motion of a charged particle is restricted to directions
perpendicular to the magnetic field and this leads to a dimensional reduction
($d\rightarrow d-2$) in the dynamics of the fermion pairing. Different methods
can be used to study this problem. Among them, the Schwinger proper time
method~\cite{Schwinger:nm} allows to obtain the exact expression for the
fermion propagator $S(x,y)$ in a constant magnetic field and evaluate the
condensate in a limiting procedure,
$\langle0|\bar{\psi}\psi|0\rangle=-\underset{x\rightarrow y}{\lim}$ Tr
$S(x,y).$ In an alternative approach, one can develop the propagator in terms
of Landau poles~\cite{Chodos:1990vv} and obtain the condensate as the
contribution coming from the lowest Landau level (LLL). In this sense, the LLL
plays a role similar to the Fermi surface in the Bardeen-Cooper-Schrieffer
(BCS) superconductivity theory. The drawback of the above methods is the fact
that they do not reveal explicitly the vacuum structure in terms of fermion
(antifermion) creation and annihilation operators. On the other hand, not only
calculations are simpler in the operator formalism but also the interpretation
of the pairing structure as well as its nonperturbative nature are more
transparent~\cite{Jona-Lasinio:1999sn}. In this paper we adhere to the spirit
of Ref.~\cite{Jona-Lasinio:1999sn} and develop a general operator formalism
necessary to study the vacuum structure and fermion condensation in systems
unyielding to the usual mean-field approach. In particular, we study the
mechanism of fermion condensation in models where, besides an external magnetic
field, a strong local or nonlocal confining interaction is also present.

The paper is organized as follows: in section \ref{sec:approach} we introduce
the formalism, based on the use of successive Valatin-Bogoliubov canonical
transformations. In section~\ref{sec:free} we exemplify its use by working out
the complete diagonalization of the Hamiltonian of a free fermion in the
presence of a constant magnetic field for both the (2+1)- and (3+1)-dimensional
cases. Section~\ref{sec:NJL} is devoted to test our formalism against the
well-known results of the local Nambu-Jona-Lasinio (NJL) model~\cite{Nambu:tp},
setting the stage for the new results concerning nonlocal kernels in the
presence of magnetic fields. In particular, we show how to translate the
formulae and results obtained within the present formalism to the ones usually
obtained through the use of mean-field
techniques~\cite{Klimenko:1991he,Gusynin:va}. This translation is a validation
of our formalism. The new and far more difficult case of a nonlocal fermion
kernel with a constant magnetic field is dealt in section \ref{sec:harmonic}.
For simplicity, we have chosen a harmonic confinement. Within the present
formalism other fermion kernels, notably the linear confinement kernel, could
be considered as well. However, the linear kernel, which involves
integro-differential mass-gap equations versus the simpler differential
equations proper of the harmonic confinement, puts, for the purpose of this
paper, quite unessential calculation complexities. The end of this section is
devoted to show that this harmonic confining, extended NJL model, already
behaves, in the presence of a magnetic field, quite differently from the
local-kernel NJL model. Our conclusions are presented in section
\ref{sec:conclusion}.

\section{The approach}
\label{sec:approach}

The Lagrangian of a relativistic fermion in an external field $A_{\mu}$ has the
standard form
\begin{equation} \label{Lmag}
\mathcal{L}= \bar{\psi}(x)\left[ i \gamma^\mu D_\mu -m \right] \psi(x)\,,
\end{equation}
where $D_{\mu}=\partial_{\mu}+ieA_{\mu}$ is the covariant
derivative.\footnote{We recall that in 2+1 dimensions there are two
inequivalent representations of the Dirac algebra, described by the matrices
$\tilde{\gamma}^{\mu}$ and $-\tilde{\gamma}^{\mu}$, with
$\tilde{\gamma}^{0}=\sigma_{3}\ ,\ \tilde{\gamma}^{1}=i\sigma_{1} \ ,\
\tilde{\gamma}^{2}=i\sigma_{2}$; $\sigma_{i}$ are the Pauli matrices. The
corresponding chiral version of the problem can then be formulated using the
4-dimensional spinor representation $\gamma^{\mu}= \text{diag}
\,(\tilde{\gamma}^{\mu},-\tilde{\gamma}^{\mu})\,.$} Different gauges can be
used to solve the Dirac equation in a constant external magnetic field. In what
follows we choose the Landau gauge $A_{\mu}=-B y\, \delta_{\mu 1}$\,, where $B$
is the magnetic field strength. In 2+1 dimensions, the presence of a
homogeneous magnetic field breaks polar symmetry explicitly, while in 3+1
dimensions, the spherical symmetry is broken. As is well known, this fact leads
to the use of Landau levels as the appropriate basis for the remaining
symmetry. The energy spectrum is given by $E^2=m^{2}+p_z^2+(2n+1-s)|eB|$, where
$s=\pm 1$ defines the spin orientation and $p_z$ is the momentum along the
$z$-direction\footnote{In 2+1 dimensions, $p_z=0$ and $s=+1$ ($s=-1$) for $E>0$
$(E<0)$.}. The discrete number $n=0,1,2,\ldots$ describes the Landau levels.

The Hamiltonian of the problem is given by
\begin{align} \label{H2D}
H = \int d^D x\, \psi^\dag(\pmb{x})\left(\beta m + \pmb{\alpha} \cdot \pmb{p} -
e\, \pmb{\alpha} \cdot \pmb{A}\right)\psi(\pmb{x}) \ ,
\end{align}
where $D$ is the space dimension and we use the standard notation
$\beta=\gamma^0$ and $\pmb{\alpha}=\gamma^0 \pmb{\gamma}$. In order to
construct the Dirac field in a magnetic field we start from the plane wave
decomposition of the free field,
\begin{align} \label{psican}
\psi(\pmb{x}) & =\sum\limits_{\pmb{p},\;s}\frac{1}{\sqrt{V}} \left\{
u_s(\pmb{p})\ a_{s \pmb{p}}+ v_s(\pmb{p})\ b_{s -\pmb{p}}^\dag \right\} e^{i
\pmb{p} \cdot \pmb{x}}  ,
\end{align}
where $V$ is the volume. Clearly, in 2+1 dimensions the spin quantum number $s$
has no meaning and it should be omitted in all the corresponding formulae. The
operators $a_{s \pmb{p}}$ and $b_{s \pmb{p}}$ satisfy the usual anticommutation
relations. The $u$ and $v$ spinors are the solutions of the Dirac equation for
positive and negative energies, respectively.

Since the quantum field of a particle in a magnetic field is usually developed
in a basis indexed by the Landau levels $n$, a change of basis is needed in
order to relate it with the plane wave solution. To achieve this, we first
perform the canonical transformation which relates the spinors
$u_{s'}(\pmb{p})\,, v_{s'}(\pmb{p})$ with the momentum-independent
($\pmb{p}=0$) spinors $\tilde{u}_s\,, \tilde{v}_s$,
\begin{align} \label{Otransf}
\left(
\begin{array}{c}
\tilde{a}_{s \pmb{p}} \\
\tilde{b}^\dag_{s -\pmb{p}}
\end{array}
\right)= \sum_{s'} O_{s s'}(\pmb{p})\left(
\begin{array}{c}
a_{s' \pmb{p}} \\
b^\dag_{s' -\pmb{p}}
\end{array}
\right) \,, \quad \left(
\begin{array}{c}
\tilde{u}_s \\
\tilde{v}_s
\end{array}
\right) = \sum_{s'} O_{s s'}^*(\pmb{p}) \left(
\begin{array}{c}
u_{s'}(\pmb{p}) \\
v_{s'}(\pmb{p})
\end{array}
\right)\,,
\end{align}
where $O_{s s'}(\pmb{p})$ is the rotation matrix. The resulting field maintains
the canonical structure (\ref{psican}). Next, in order to index the spinors and
operators by the Landau levels, we expand the field in a complete set of
Hermite polynomials $H_n$ through the identity
\begin{align} \label{chbasis}
e^{i p_y y}=e^{-i \ell^2 p_x p_y}\sqrt{2\pi}\sum_{n=0}^\infty\, i^n
\omega_n(\xi)\,\omega_n(\ell p_y)\,,
\end{align}
where
\begin{align}\label{omegan}
\omega_n(x)=(2^n n! \sqrt{\pi})^{-1/2}e^{-x^2/2}H_n(x)
\end{align}
and
\begin{align}
\xi=y/\ell+\ell p_x\,,
\end{align}
$ \ell=|eB|^{-1/2}$ is the magnetic length. The corresponding change of basis
is given by
\begin{align} \label{Ttransf}
\left(
\begin{array}{c}
\hat{a}_{s n \bar{p}} \\
\hat{b}^\dag_{s n -\bar{p}}
\end{array}
\right) = \sum_{s',\; p_y} \tilde{T}_{n s s'}(\ell p_y) \left(
\begin{array}{c}
\tilde{a}_{s' \pmb{p}} \\
\tilde{b}^\dag_{s' -\pmb{p}}
\end{array}
\right) ,\, \quad \left(
\begin{array}{c}
\hat{u}_{s n}(\xi) \\
\hat{v}_{s n}(\xi)
\end{array}
\right)= \sum_{s'} T_{n s s'}(\xi) \left(
\begin{array}{c}
\tilde{u}_{s'} \\
\tilde{v}_{s'}
\end{array}
\right)\,,
\end{align}
where $\bar{p}=(p_x,p_z)$ in 3+1 dimensions and $\bar{p}=p_x$ in 2+1
dimensions, and the operators satisfy the usual anticommutation relations
$\{\hat{a}^\dag_{s n \bar{p}}\,,\hat{a}_{s' n' \bar{p}'}\} = \{\hat{b}^\dag_{s
n \bar{p}}\,,\hat{b}_{s' n' \bar{p}'}\} =\delta_{s s'}\, \delta_{n n'}\,
\delta_{\bar{p} \bar{p}'} \,$. The matrices $\tilde{T }_n(\ell p_y)$ and
$T_n(\xi)$ can be obtained after a simple algebra.

The last step consists in obtaining the mass gap equations which lead to the
diagonalization of the Hamiltonian. There are several approaches one can adopt
to obtain the mass gap equation. It can be derived as the condition for the
vacuum energy to be a minimum~\cite{LeYaouanc:1983iy,Alkofer:1988tc}, or in the
form of a Dyson equation for the fermion propagator~\cite{Alkofer:1988tc}, or
as a Ward identity~\cite{Adler:1984ri}. Here we shall derived it by imposing
the vanishing of the anomalous (nondiagonal) terms in the Hamiltonian,
corresponding to the direct creation (or annihilation) of a fermion-antifermion
pair~\cite{Schrieffer:1964,Bicudo:sh}. Aiming at this, we perform a second
canonical transformation defined as
\begin{align}\label{Rtransf}
\left(
\begin{array}{c}
a_{s n \bar{p}} \\
b^\dag_{s n -\bar{p}}
\end{array}
\right) = \sum_{s'} R_{n ss'}(p_z)\left(
\begin{array}{c}
\hat{a}_{s' n \bar{p}} \\
\hat{b}^\dag_{s' n -\bar{p}}
\end{array}
\right) , \, \quad \left(
\begin{array}{c}
u_{s n}(\xi) \\
v_{s n}(\xi)
\end{array}
\right)= \sum_{s'} R^*_{n ss'}(p_z) \left(
\begin{array}{c}
\hat{u}_{s' n}(\xi) \\
\hat{v}_{s' n}(\xi)
\end{array}
\right)\,,
\end{align}
where the rotation matrix $R_{n ss'}(p_z)$ is obtained by requiring the
anomalous terms in the Hamiltonian to vanish.

After this diagonalization procedure, the vacuum state in the magnetic field
can be easily constructed. Indeed, each Valatin-Bogoliubov transformation on
the operators and spinors corresponds to a transformation of the vacuum such
that the latter is annihilated by the new operators. In the next sections we
give explicit realizations of this construction.

The approach presented above can also be extended to the study of systems
having, besides an external magnetic field, a quartic fermion interaction of
the form
\begin{align}\label{hintera}
H_{int}=\int d^D x \int d^D y \, \bar{\psi}(\pmb{x}) \Gamma \psi(\pmb{x}) \,
V(|\pmb{x}-\pmb{y}|) \, \bar{\psi}(\pmb{y}) \Gamma \psi(\pmb{y}) \, ,
\end{align}
where $\Gamma$ denotes the vertex (Dirac, color) structure, and $V$ is an
attractive potential. This type of kernels has been extensively used in
hadronic phenomenology~\cite{Bicudo:sh,hadron}. In this case, we write the
total Hamiltonian in normal-ordered form with respect to the new vacuum. Using
the Wick contraction technique, the Hamiltonian can be decomposed in three
terms,
\begin{align} \label{Hdecomp}
H &=\ :\!H_0\!: + :\!H_2\!: + :\!H_4\!:\,,\\ \nonumber \\
:\!H_0\!: &\propto\, \wick{1}{<1\psi_\alpha^\dag K
>1\psi_\alpha} +
\wick{21}{<2\psi_\alpha^\dag \Gamma <1\psi_\beta V
>1\psi_\beta^\dag \Gamma >2\psi_\alpha} +
\wick{11}{<1\psi_\alpha^\dag \Gamma >1\psi_\alpha V <+\psi_\beta^\dag \Gamma
>+\psi_\beta}\,, \nonumber\\
:\!H_2\!: &\propto\,  \psi_\alpha^\dag K \psi_\beta +
 \wick{1}{<1\psi_\alpha^\dag \Gamma
\psi_\beta V \psi_\gamma^\dag \Gamma >1\psi_\alpha} +
\wick{1}{<1\psi_\alpha^\dag \Gamma >1\psi_\alpha V \psi_\beta^\dag \Gamma
\psi_\gamma} \nonumber\\
& \hspace{3cm}+ \wick{1}{\psi_\alpha^\dag \Gamma <1\psi_\beta V
>1\psi_\beta^\dag \Gamma \psi_\gamma} + \wick{1}{\psi_\alpha^\dag \Gamma
\psi_\beta V
<1\psi_\gamma^\dag \Gamma >1\psi_\gamma}\,, \nonumber\\
:\!H_4\!: &\propto\, \psi_\alpha^\dag \Gamma \psi_\beta V \psi_\gamma^\dag
\Gamma \psi_\delta\,, \nonumber
\end{align}
containing 0, 2 and 4 creation and/or annihilation operators, respectively. In
the presence of a magnetic field, $K = \beta m + \pmb{\alpha} \cdot \pmb{p} -
e\, \pmb{\alpha} \cdot \pmb{A}$. The Greek indices refer to the two different
Dirac fields, $\psi_1$ and $\psi_2$, corresponding to two inequivalent
representations of the Dirac algebra in the (2+1)-dimensional case. In 3+1
dimensions, such indexing is not necessary.

\begin{figure}[t]
\begin{center}
\includegraphics{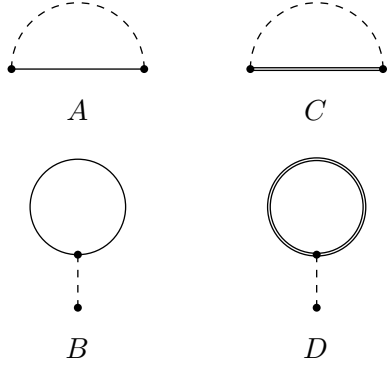}
\end{center}
\caption{Diagrams contributing to $:\!H_2\!:\,$. The dashed lines represent the
interaction potential $V$ and the single (double) lines correspond to the
fermion propagator of the field $\psi_1\, (\psi_2)\,$ in the case of 2+1
dimensions.} \label{diagrams}
\end{figure}

The contribution to $:\!H_2\!:$ coming from $H_{int}$ can be diagrammatically
expressed in 2+1 dimensions as
\begin{align}\label{H2in2D}
 \psi^\dag_1 \Gamma \psi_1 V \psi^\dag_1 \Gamma \psi_1 + \psi^\dag_2 \Gamma
\psi_2 V \psi^\dag_2 \Gamma \psi_2 &  \longrightarrow \psi^\dag_1 \left[ A-B
\right]\,\psi_1 + \psi^\dag_2 \left[ C-D \right]\,\psi_2 \,, \nonumber\\
 \psi^\dag_1 \Gamma \psi_1 V \psi^\dag_2 \Gamma \psi_2 + \psi^\dag_2 \Gamma
\psi_2 V \psi^\dag_1 \Gamma \psi_1 & \longrightarrow \psi^\dag_1 \left[ D
\right]\,\psi_1 + \psi^\dag_2 \left[ B \right]\,\psi_2 \,,\\
 \psi^\dag_1 \Gamma \psi_2 V \psi^\dag_2 \Gamma \psi_1 + \psi^\dag_2 \Gamma
\psi_1 V \psi^\dag_1 \Gamma \psi_2 & \longrightarrow \psi^\dag_1 \left[ C
\right]\,\psi_1 + \psi^\dag_2 \left[ A \right]\,\psi_2 \,, \nonumber
\end{align}
where $A,\,B,\,C$ and $D$ correspond to the diagrams presented in
Fig.~\ref{diagrams}. On the other hand, in 3+1 dimensions the contribution is
simply
\begin{align} \label{H2in3D}
\psi^\dag \Gamma \psi V \psi^\dag \Gamma \psi \longrightarrow \psi^\dag \left[
A-B\right]\,\psi \,.
\end{align}
Once the terms contributing to the quadratic part $:\!H_2\!:$ have been
identified, its diagonalization proceeds through a sequence of
Valatin-Bogoliubov transformations, as given in Eqs.~(\ref{Otransf}),
(\ref{Ttransf}) and (\ref{Rtransf}).

We note that, in the calculation of the diagrams of Fig.~\ref{diagrams}, the
fermion propagators associated with $\psi_i\, (i=1,2)$ in 2+1 dimensions take
the explicit form
\begin{align}
S^{(i)}(\pmb{x}, \pmb{x'})= \frac{1}{\ell L_x} \sum\limits_{n,\;p_x} \left[
u_n^{(i)}(\xi_{y})\,{u_n^{(i)}}^\dag(\xi_{y'}) -
v_n^{(i)}(\xi_{y})\,{v_n^{(i)}}^\dag(\xi_{y'})\right] e^{i(x-x')p_x}\,,
\label{prop2D}
\end{align}
while in the (3+1)-dimensional case the propagator is given by
\begin{align} \label{prop3D}
S(\pmb{x}, \pmb{x'}) = \frac{1}{\ell L_x L_z} \sum\limits_{n, \bar{p}, s}\left[
u_{sn}(\xi_{y}) u_{sn}^\dag(\xi_{y'}) - v_{sn}(\xi_{y})
v_{sn}^\dag(\xi_{y'})\right] e^{i((x-x')p_x+(z-z')p_z)}\,.
\end{align}

\section{Free fermions in a magnetic field}
\label{sec:free}

As a first example of the applicability of our approach, we consider the
problem of a fermion in 2+1 dimensions in the presence of a constant magnetic
field. In this case, the rotation matrix $O(\pmb{p})$ corresponding to the
first canonical transformation (\ref{Otransf}) is simply given by
\begin{align}
O(\pmb{p}) = \left(
\begin{array}{cc}
\cos \phi & -\sin \phi\ (\hat{p}_y + i \hat{p}_x) \\
\sin \phi\ (\hat{p}_y - i \hat{p}_x) & \cos \phi
\end{array}
\right) , \,\quad \hat{\pmb{p}} =\frac{\pmb{p}}{|\pmb{p}|}\,,\,\quad \tan 2\phi
= \frac{|\pmb{p}|}{m}\,.
\end{align}
The vacuum associated to the operators $\tilde{a}_{\pmb{p}}$ and
$\tilde{b}_{\pmb{p}}$ has the form
\begin{align}
|\tilde{0}\rangle = \prod_{\pmb{p}} (\cos \phi + \sin \phi \, a^\dag_{\pmb{p}}
b^\dag_{-\pmb{p}})|0 \rangle\,,
\end{align}
so that $ \tilde{a}_{\pmb{p}} |\tilde{0} \rangle = 0 ,\ \tilde{b}_{\pmb{p}}
|\tilde{0}\rangle =0 $.

The second canonical transformation defined in Eq.~(\ref{Rtransf}) is performed
with the rotation
\begin{align} \label{Rtheta}
R_{n} = \left(
\begin{array}{cc}
\cos \theta_n & -\sin \theta_n \\
\sin \theta_n & \cos \theta_n
\end{array}
\right)\,,
\end{align}
where the angles $\theta_n$ are determined by imposing the vanishing of the
anomalous terms in the Hamiltonian. A simple algebraic computation yields the
following mass gap equations,
\begin{align} \label{massgap2D}
\left\{
\begin{array}{ll}
m \sin 2\theta_0 = 0\ , &\quad n= 0\ ,\\
m \sin 2\theta_n -  \ell^{-1}\,\sqrt{2n}\, \cos 2\theta_n  = 0 \ , &\quad n>0\
,
\end{array}
\right.
\end{align}
which for any value of $n$ have the solution
\begin{align}
\tan 2\theta_n=\frac{\sqrt{2n}}{m\ell}\ .
\end{align}

It remains to construct the vacuum state in the magnetic field $|0
\rangle_B\,$, which is annihilated by the operators $a_{n p_x}$ and $b_{n
p_x}\,$. We find
\begin{align}
|0 \rangle_B & = \prod_{n,\;p_x}(\cos \theta_n  + \sin \theta_n \,
\hat{a}^\dag_{n p_x} \hat{b}^\dag_{n -p_x})|\tilde{0}\rangle\ .
\end{align}

At this point, it is worth commenting on the question of dynamical symmetry
breaking in the presence of the magnetic
field~\cite{Gusynin:va,Jona-Lasinio:1999sn}. This problem is connected with the
appearance of a fermion condensate ${_B}\langle 0 | \bar{\psi} \psi | 0
\rangle_B \neq 0$ in the limit where the mass parameter $m$ goes to zero.
Writing the field $\psi$ in terms of the new canonical operators one can easily
obtain
\begin{align} \label{qqcond2}
\lim_{m\rightarrow 0}{_B} \langle 0 | \bar{\psi} \psi | 0 \rangle_B & = -
\lim_{m\rightarrow 0} \frac{1}{2 \pi \ell^2}\left(\cos 2\theta_0 + 2
\sum_{n=1}^\infty
\cos 2\theta_n \right) \nonumber\\
& = - \lim_{m\rightarrow 0} \frac{1}{2\pi \ell^2}\left(1 + 2 \sum_{n=1}^\infty
\frac{m}{\sqrt{m^2 + 2n/\ell^2}} \right) = - \frac{|eB|}{2\pi} \ ,
\end{align}
i.e. the spontaneous breaking of the original $U(2)$ symmetry of the Lagrangian
(\ref{Lmag}) occurs even in the absence of any additional interaction between
fermions~\cite{Gusynin:va}. This result is specific to the (2+1)-dimensional
Dirac theory in an external magnetic field.

Let us now consider the problem in 3+1 dimensions. The rotation matrix $O_{s
s'}(\pmb{p})$ in Eq.~(\ref{Otransf}) is given in this case by
\begin{align}
O_{s s'}(\pmb{p}) &= \left(
\begin{array}{cc}
\cos \phi\ \delta_{s s'} & -\sin \phi\ M_{s s'} \\
\sin \phi\ M_{s s'}^\dag & \cos \phi\ \delta_{s s'}
\end{array}
\right) ,\,\quad  \tan 2\phi = \frac{|\pmb{p}|}{m}\,, \,\quad M=-(
\hat{\pmb{p}}\cdot\pmb{\sigma} )(i \sigma_2)\,.
\end{align}

The new vacuum $|\tilde{0}\rangle$ can be generated from the
$|0\rangle$ vacuum following the standard procedure and we find
\begin{align}
|\tilde{0}\rangle &= \prod_{\pmb{p}} \left(\frac{1+\cos 2\phi}{2} + \frac{\sin
2\phi}{2} \, C^\dag_p + \frac{1-\cos 2\phi}{4} \, {C^\dag_p}^2 \right)|0
\rangle, \nonumber\\
C^\dag_p &= \left(
\begin{array}{cc}
a^\dag_{\uparrow p} & a^\dag_{\downarrow p}
\end{array} \right) M \left(
\begin{array}{c}
b^\dag_{\uparrow -p} \\
b^\dag_{\downarrow -p}
\end{array} \right)\,,
\end{align}
which obviously satisfies $\tilde{a}_{s \pmb{p}} |\tilde{0} \rangle = 0, \
\tilde{b}_{s \pmb{p}} |\tilde{0}\rangle =0$.

Next, we define the rotation matrix $R_{n s s'}(p_z)$ of Eq.~(\ref{Rtransf}) as
\begin{align} \label{Rthetaphi}
R_{n s s'}(p_z) = \left(
\begin{array}{cc}
\cos \theta_n\, \delta_{s s'} & -\sin \theta_n {M_n}_{s s'} \\
\sin \theta_n {M_n^\dag}_{s s'} & \cos \theta_n\, \delta_{s s'}
\end{array}
\right) , \,\quad M_n =\left(
\begin{array}{cc}
-\sin \varphi_n & \cos \varphi_n \\
\cos \varphi_n & \sin \varphi_n
\end{array}
\right),
\end{align}
with $M_n$ chosen so that $M^\dag_n M_n =\openone$ and $\varphi_0=0$. Once
more, the angles $\theta_n$ and $\varphi_n$ are determined by requiring the
anomalous terms in the Hamiltonian to vanish. This leads to the following mass
gap equations,
\begin{align}
\left\{
\begin{array}{ll}
m \sin 2\theta_0 - p_z \cos 2\theta_0 = 0\,, & n= 0\ ,\medskip\\
m \sin 2\theta_n M_n - \sqrt{p_z^2+2n/\ell^2}\, \left[\cos^2 \theta_n
\tilde{M}_n -\sin^2 \theta_n M_n \tilde{M}_n M_n\right] = 0\,, & n>0\ ,
\end{array}
\right.
\end{align}
where
\begin{equation} \label{Mtilde}
\tilde{M}_n = \frac{1}{\sqrt{p_z^2 + 2n/\ell^2}} \left(
\begin{array}{cc}
-\sqrt{2n}/\ell & p_z \\
p_z & \sqrt{2n}/\ell
\end{array} \right)
\,.
\end{equation}
The solution of the above equations is given for any value of $n$ by
\begin{align}
\tan\varphi_n =\frac{\sqrt{2n}}{p_z \ell}\,\,, \quad \tan 2\theta_n =
\frac{\sqrt{p_z^2+2n/\ell^2}}{m}\,\,.
\end{align}

The vacuum state in the magnetic field reads in this case
\begin{align}
|0\rangle_B &= \prod_{n,\;\bar{p}} \left(\frac{1+\cos 2\theta_n}{2} +
\frac{\sin 2\theta_n}{2} \, C^\dag_n + \frac{1-\cos 2\theta_n}{4} \,
{C^\dag_n}^2
\right)|\tilde{0} \rangle\ , \nonumber\\
C^\dag_n &= \left(
\begin{array}{cc}
\hat{a}^\dag_{\uparrow n \bar{p}} & \hat{a}^\dag_{\downarrow n
\bar{p}}
\end{array} \right) M_n \left(
\begin{array}{c}
\hat{b}^\dag_{\uparrow n -\bar{p}} \\
\hat{b}^\dag_{\downarrow n -\bar{p}}
\end{array} \right)\,.
\end{align}

Let us now evaluate the fermion condensate ${_B}\langle 0 | \bar{\psi} \psi | 0
\rangle_B$. In the limit when $m$ goes to zero we obtain
\begin{align} \label{qqcond3}
\lim_{m\rightarrow 0}{_B}\langle 0 | \bar{\psi} \psi | &0 \rangle_B = -
\lim_{m\rightarrow 0} \frac{1}{2 \pi \ell^2} \int\frac{d p_z}{2\pi} \left(\cos
2\theta_0 + 2
\sum_{n=1}^\infty \cos 2\theta_n \right) \nonumber \\
&= - \lim_{m\rightarrow 0} \frac{1}{2\pi^2 \ell^2} \int_0^\Lambda d p_z
\left(\frac{m}{\sqrt{m^2+p_z^2}} + 2
\sum_{n=1}^\infty \frac{m}{\sqrt{m^2 + p_z^2 + 2n/\ell^2}} \right) \nonumber \\
&= -\lim_{m\rightarrow 0} \frac{|eB|}{4\pi^2}\, m \left[\ln
\frac{\Lambda^2}{m^2}+\mathcal{O}(m^0)\right] \rightarrow 0
 \ ,
\end{align}
where an ultraviolet cutoff $\Lambda$ has been introduced in order to
regularize the integral. In contrast to the (2+1)-dimensional case, where a
nonvanishing condensate appears in the limit when $m\rightarrow 0$ (see Eq.
(\ref{qqcond2})), it is seen from Eq.~(\ref{qqcond3}) that this quantity
vanishes in the (3+1)-dimensional case. Nevertheless, Eq.~(\ref{qqcond3})
indicates that a condensate in 3+1 dimensions could in principle be generated
in the presence of a small dynamical mass.

\section{A test example: the NJL local kernel}
\label{sec:NJL}

At this stage it is convenient to test our formalism against the well-known
results of the NJL model. This test, which is nontrivial, will then act as a
validation criteria for the present formalism.

The Lagrangian density for the original NJL model in an external
field reads
\begin{align} \label{LNJL}
\mathcal{L} = \bar{\psi}(\pmb{x}) & \left[ i \gamma^\mu D_\mu -m \right]
\psi(\pmb{x}) + G \, \left[(\bar{\psi}(\pmb{x}) \psi(\pmb{x}))^2 +
(\bar{\psi}(\pmb{x}) i \gamma_5 \psi(\pmb{x}))^2\right]\, .
\end{align}

Let us start by discussing the (2+1)-dimensional NJL model. We recall that in
2+1 dimensions the $\gamma_5$ matrix in the chiral representation is defined as
\begin{align}
\gamma^{5}=i\left(
\begin{array}[c]{cc}
0 & -\openone\\
\openone & 0
\end{array}
\right)\,,
\end{align}
where $\openone$ is the $2\times2$ unit matrix. The contribution of $H_{int}$
to $:\!H_2\!:$ is easily obtained from Eq.~(\ref{H2in2D}),
\begin{align}\label{2plus1h2}
\psi_1^\dag \left[ S^{(1)}+S^{(2)} - \mbox{Tr} \left(S^{(1)}-S^{(2)}\right)
\right] \psi_1 + (1\leftrightarrow 2)\,,
\end{align}
where the fermion propagators are given by (cf. Eq.~(\ref{prop2D}))
\begin{align}
S^{(i)}=\frac{1}{\ell L_x} \sum\limits_{n,\;p_x} \left[
u_n^{(i)}(\xi){u_n^{(i)}}^\dag(\xi)-
v_n^{(i)}(\xi){v_n^{(i)}}^\dag(\xi)\right]\,.
\end{align}

Following the procedure of section \ref{sec:approach}, the diagonalization of
the Hamiltonian $:\!H_2\!:$ through Valatin-Bogoliubov transformations yields
the following mass gap equations:
\begin{align}
\left\{
\begin{array}{ll}
(m + \rho)\,\sin 2\theta_0 = 0\ , &\quad n= 0\, ,\\
(m+\rho) \sin 2\theta_n - \ell^{-1}\,\sqrt{2n}\, \cos 2\theta_n = 0 \ , &\quad
n>0\, ,
\end{array}
\right.
\end{align}
where
\begin{align} \label{rho2D}
\rho=\frac{G}{\pi\ell^2} \left(\cos 2 \theta_0 + 2 \sum_{n=1}^\infty \cos
2\theta_{n}\right)\,.
\end{align}
The solution of these equations is
\begin{align}
\tan 2\theta_n=\frac{\sqrt{2n}}{(m+\rho)\ell} \,,
\end{align}
which when substituted back into Eq.~(\ref{rho2D}) yields
\begin{align} \label{massgapNJL2D}
\frac{\pi \ell^2}{G}\rho = 1+2\sum_{n=1}^\infty
\frac{m+\rho}{\sqrt{2n/\ell^2+(m+\rho)^2}}\, .
\end{align}
Using the identity~\cite{Gradshteyn}
\begin{align} \label{ident}
2\sum_{n=0}^\infty \frac{1}{(2n+\beta)^\mu}=\beta^{-\mu}+ \frac{1}{\Gamma(\mu)}
\int_0^\infty ds s^{\mu-1} e^{-\beta s} \coth s \ ,
\end{align}
we can recast Eq.~(\ref{massgapNJL2D}) in the form
\begin{align}
\frac{\pi^{3/2} \ell}{G}\rho = (m+\rho)\int^\infty_0 \frac{d s}{s^{1/2}}
e^{-(m+\rho)^2 \ell^2 s} \coth s\, ,
\end{align}
which reproduces the result obtained in leading order of the $1/N$
expansion~\cite{Gusynin:va}.

In the weak coupling regime, i.e. when $G \ll \ell\,$, one expects the lowest
Landau level to dominate. In this case, Eq.~({\ref{massgapNJL2D}) implies
\begin{align}
\rho=\frac{G}{\pi\ell^2}=G\frac{|eB|}{\pi}\,.
\end{align}
Recalling that $\langle 0 | \bar{\psi} \psi |0 \rangle = -\rho/(2G)\,$, we find
$\langle 0 | \bar{\psi} \psi |0 \rangle = -|eB|/(2\pi)$. We notice that this
result coincides with the value of the condensate previously obtained in
Eq.~(\ref{qqcond2}), in the absence of the NJL interaction.

Let us now consider the (3+1)-dimensional NJL model. In this case, the
contribution to $:\!H_2\!:$ is calculated from Eq.~(\ref{H2in3D}) and one
obtains
\begin{align}
 \begin{split}
\psi^\dag \left[ \gamma_0 S \gamma_0 - \gamma_0 \mbox{Tr} \left(S
\gamma_0\right) - \gamma_0\gamma_5 S \gamma_0\gamma_5 + \gamma_0\gamma_5
\mbox{Tr} \left(S \gamma_0\gamma_5\right) \right] \psi \,,
 \end{split}
\end{align}
where, according to Eq.~(\ref{prop3D}), the fermion propagator is
\begin{align}
S=\sum\limits_{n,\;\bar{p}}\frac{1}{\ell L_x L_z} \sum_{s}\left[
u_{sn}(\xi)u_{sn}^\dag(\xi)- v_{sn}(\xi)v_{sn}^\dag(\xi)\right]\,.
\end{align}

Once again, $:\!H_2\!:$ is diagonalized by successive Valatin-Bogoliubov
transformations. For the sake of comparison with the results obtained using the
$1/N$-expansion technique, we shall assume that pions do not condensate. In
this case, we end up with the following mass gap equations:
\begin{align}
\left\{
\begin{array}{ll}
\left(m + \rho \right) \sin 2\theta_0 - \left (p_z + \kappa\right)
\cos 2\theta_0 = 0\,, &\, n= 0\,,\medskip\\
(m+\rho) \sin 2\theta_n M_n - \sqrt{p_z^2+2n/\ell^2} \left(\cos^2 \theta_n
\tilde{M}_n - \sin^2 \theta_n M_n \tilde{M}_n M_n\right) \\
\hspace{2cm} - \kappa \left[ \cos^2 \theta_n \sigma_1 + \sin^2 \theta_n
(\sin2\varphi_n \sigma_3 - \cos2\varphi_n \sigma_1)\right] = 0 \,, &\, n>0\,,
\end{array}
\right.
\end{align}
where $M_n$ and $\tilde{M}_n$ are defined in Eqs.~(\ref{Rthetaphi}) and
(\ref{Mtilde}), respectively;
\begin{align} \label{rho3D}
&\rho = \frac{G}{\pi \ell^2}\int \frac{d p_z}{2\pi} \left( \cos
2\theta_0 +2 \sum_{n=1}^\infty \cos
2\theta_n\right)\,, \nonumber\\
&\kappa = \frac{G}{\pi \ell^2}\int \frac{d p_z}{2\pi} \left( \sin 2\theta_0 +2
\sum_{n=1}^\infty \sin 2\theta_n \cos \varphi_n\right)\,.
\end{align}

The solution of these equations is given by
\begin{align}
\tan \varphi_n =\frac{\sqrt{2n}}{(p_z + \kappa) \ell}\,, \quad \tan 2\theta_n
=\frac{\sqrt{(p_z + \kappa)^2+2n/\ell^2}}{m+\rho }\,.
\end{align}
Thus, from Eq.~(\ref{rho3D}) we find
\begin{align} \label{NJL3final}
\frac{2 \pi^2 \ell^2}{G}\rho &= \int d p_z \left(
\frac{m+\rho}{\sqrt{p_z^2+(m+\rho)^2}} + 2\sum_{n=1}^\infty
\frac{m+\rho}{\sqrt{2n/\ell^2 +p_z^2+(m+\rho)^2}} \right)\,,
\end{align}
which can be recast in the integral form
\begin{align}
\frac{2\pi^2 \ell^2}{G}\rho =(m+\rho)\int^\infty_0 \frac{d s}{s}
e^{-(m+\rho)^2\ell^2 s} \coth s\,,
\end{align}
through the identity (\ref{ident}).

In the weak coupling regime, when the LLL dominance approximation remains
valid, from Eq.~(\ref{NJL3final}) we obtain in the limit $m \rightarrow
0$,
\begin{align}\label{cond3dLLL}
    \frac{2 \pi^2 \ell^2}{G}\rho = \rho \ln
    \frac{\Lambda^2}{\rho^2}\,,
\end{align}
where $\Lambda$ is an ultraviolet cutoff which regularizes the integral in the
momentum $p_z$. This result is consistent with the one previously found in the
absence of the NJL interaction term [cf. Eq.~(\ref{qqcond3})] and agrees with
the results obtained by the use of other approaches~\cite{Gusynin:va}. We also
notice that Eq.~(\ref{cond3dLLL}) gives the correct behavior near the
singularity at $G=0$,
\begin{align} \label{NJLmass}
 \rho^2=\Lambda^2 \exp \left(-\frac{2\pi^2}{G|eB|}\right)\,.
\end{align}
On the other hand, as $G$ increases the expression (\ref{NJLmass}) gets less
accurate. This is due to the fact that for larger values of $G$ all the Landau
levels give important contributions to the dynamical mass and the LLL
approximation is no longer valid.

It is worth emphasizing that in the local NJL model the nontrivial chiral
condensate is generated in the presence of an arbitrary small external magnetic
field for arbitrary small values of the coupling constant. It turns out, as we
shall see shortly, that this picture ceases to be valid when, in the presence
of a homogeneous magnetic field $B$, we go from a local to a nonlocal fermionic
kernel. In this case we cease to have a weak coupling limit no matter how small
we set the kernel strength.

\section{Nonlocal kernels: the harmonic oscillator case}
\label{sec:harmonic}

The (2+1)-dimensional case of a nonlocal quartic kernel (given in this
example, for the sake of simplicity, by a harmonic-oscillator kernel)
constitutes, in the presence of a magnetic field, the first nontrivial
application of our formalism.

The harmonic-oscillator quartic interaction is given by
\begin{align}
H_{int}=\!\int \!d^2 x\, d^2 y \, \bar{\psi}(\pmb{x}) \gamma_0
\frac{\lambda^a}{2} \psi(\pmb{x}) ~\frac{3}{4}\, K |\pmb{x}-\pmb{y}|^2~
\bar{\psi}(\pmb{y}) \gamma_0 \frac{\lambda^a}{2} \psi(\pmb{y})\, ,
\end{align}
where $K$ is the interaction constant and $\lambda^a$ are the Gell-Mann
matrices, which account for the color structure of the interaction. This
structure prevents the appearance of tadpole-like terms. Notice that $H_{int}$
has a central (polar) symmetry.

In the absence of a magnetic field, the harmonic-oscillator nonlocal NJL
Hamiltonian can be readily diagonalized through a ``2D polar-symmetric"
Valatin-Bogoliubov canonical transformation. The other limit, i.e. having a
magnetic field and no quartic fermion interaction, has been already studied in
the present paper and leads to a condensate with nonpolar, axial-like symmetry.
It is therefore clear that we cannot accommodate both symmetries
simultaneously: if we had started from a sufficiently strong magnetic field,
and progressively turned on a quartic nonlocal interaction, there will be a
value of the quartic kernel strength where the fermion condensate cannot hold
any longer and the system becomes disordered.

The contribution to the Hamiltonian term $:\!H_2\!:$ in Eq.~(\ref{H2in2D}) is
represented by the diagrams $A$ and $C$ of Fig.~\ref{diagrams},
\begin{align}
\psi_1^\dag (\pmb{x}) S^{(1)}(\pmb{x},\pmb{y}) V(|\pmb{x}-\pmb{y}|)
\psi_1(\pmb{y}) + (1\leftrightarrow 2)\,.
\end{align}
Here, $S^{(1)}(\pmb{x},\pmb{y})$ is the propagator defined in
Eq.~(\ref{prop2D}). The tadpole diagrams $B$ and $D$ do not contribute since
the $\lambda^a$ matrices are traceless. Moreover, there is neither a
contribution from diagram $C$ to the $\psi_1$ term nor a contribution from
diagram $A$ to the $\psi_2$ term, because the $\gamma_0$ vertex is block
diagonal.

Applying successive Valatin-Bogoliubov transformations and making use of the
simple relations
\begin{align}
\begin{array}{l}
(\xi - \partial_{\xi})\, \omega_n(\xi)=\sqrt{2(n+1)}\, \omega_{n+1}(\xi)\,, \\
(\xi + \partial_{\xi})\, \omega_n(\xi)=\sqrt{2n}\, \omega_{n-1}(\xi)\,,
\end{array}
\end{align}
with $\omega_n(\xi)$ defined in Eq.~(\ref{omegan}), we arrive at the following
recurrence system of mass gap equations:
\begin{align} \label{mgeosc}
\left\{
\begin{array}{ll}
\left[\dfrac{m}{2} + K\ell^2 (\cos 2\theta_0 + \cos 2\theta_1)\right] \,\sin
2\theta_0 = 0\,, &\quad n= 0\,,\medskip\\
 m \sin 2\theta_n - \sqrt{2n}/\ell \cos 2\theta_n \\
+ K \ell^2 \left[-((2n-1)\cos 2\theta_{n-1}+(2n+1)\cos
2\theta_{n+1}) \sin 2\theta_n \right.\\
\left. + 2\,\sqrt{n}(\sqrt{n-1} \sin 2\theta_{n-1} + \sqrt{n+1} \sin
2\theta_{n+1})\cos 2\theta_n \right]= 0 \,, &\quad n>0\,.
\end{array}
\right.
\end{align}
We impose the boundary condition $\theta_0=0\,$, so that the standard
expression for the $n=0$ spinor, $ u_0(\xi )=(\omega_0(\xi)\,,\,0)^T$, is
recovered. We also note that Eqs.~(\ref{mgeosc}) enforce $\theta_n \rightarrow
\pi/4\,$ as $n \rightarrow \infty\,$.

Using the second equation in (\ref{mgeosc}) we can calculate the chiral angle
$\theta_{n+1}$ from $\theta_n$ and $\theta_{n-1}$. This leaves $\theta_1$ as a
free parameter. By varying $\theta_1$, we can find the solution that minimizes
the vacuum energy,
\begin{align}
    H_0 &=-m \left(\cos 2\theta_0+2 \sum_{n=1}^{\infty}\cos
    2\theta_n\right) + K\ell^2 \sin^2 2\theta_0 -\frac{2}{\ell} \sum_{n=1}^{\infty}\sqrt{2n}
    \sin 2\theta_n\,.
\end{align}

\begin{figure}[t]
\begin{center}
\includegraphics{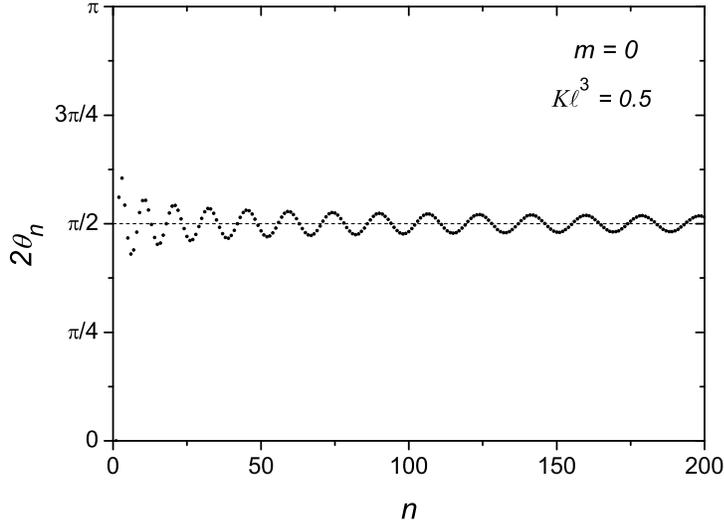}
\end{center}
\caption{Chiral angle as a function of the Landau level $n$ for $m=0$ and $K
\ell^3= 0.5$ in the presence of a harmonic-oscillator potential and a constant
magnetic field in 2+1 dimensions.} \label{thetavsn}
\end{figure}

The nonlinear mass gap equations (\ref{mgeosc}) have no obvious analytical
solutions and we have to solve them numerically. As an example, we present in
Fig.~\ref{thetavsn} their numerical solution for $m=0$ and $K \ell^3 =0.5$.
Having obtained the chiral angles we can proceed to calculate physical
quantities. The chiral condensate $\langle \bar{\psi} \psi \rangle$ was
numerically shown to converge for each value of $K \ell^3$ and, therefore, no
renormalization is needed. This is in contradistinction to the NJL local
kernel, which needs to be regularized. The results are shown in
Fig.~\ref{condensatevschi}.
\begin{figure}[t]
\begin{center}
\includegraphics{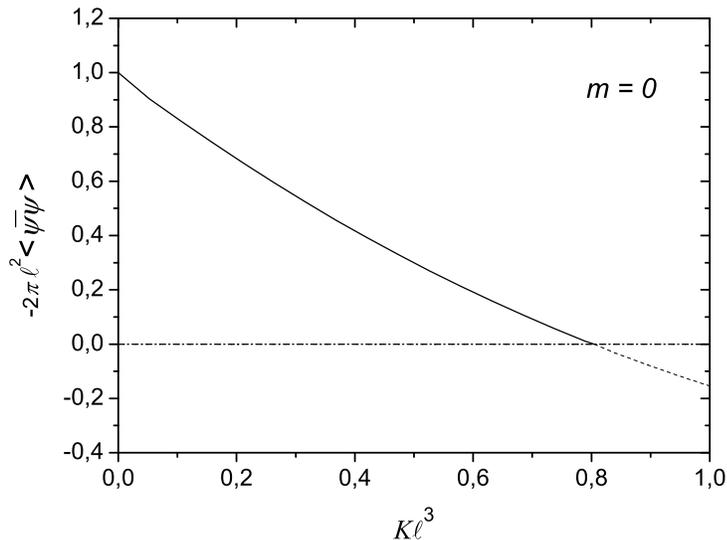}
\end{center}
\caption{Chiral condensate as a function of the dimensionless parameter $K
\ell^3$ for $m=0$, in the presence of a harmonic-oscillator potential and a
constant magnetic field in 2+1 dimensions.} \label{condensatevschi}
\end{figure}
Notice that when $K\ell^3 \rightarrow 0$ we recover the value of the condensate
previously obtained without the harmonic-oscillator interaction, i.e., $-2\pi
\ell^2 \langle 0| \bar{\psi} \psi |0\rangle =1$. When $K\ell^3\simeq 0.8$ the
chiral condensate vanishes.\footnote{Taking, for instance,
$K=(300\,\text{MeV})^3$, a value of $K\ell^3 = 0.8$ would correspond to a
magnetic field strength of $B = 1.5\times 10^{18}$~G.} This signals the onset
of a phase transition due to the presence of both the magnetic field and the
fermionic quartic kernel (in this case, a confining harmonic-oscillator
kernel).

\begin{figure}[t]
\begin{center}
\includegraphics{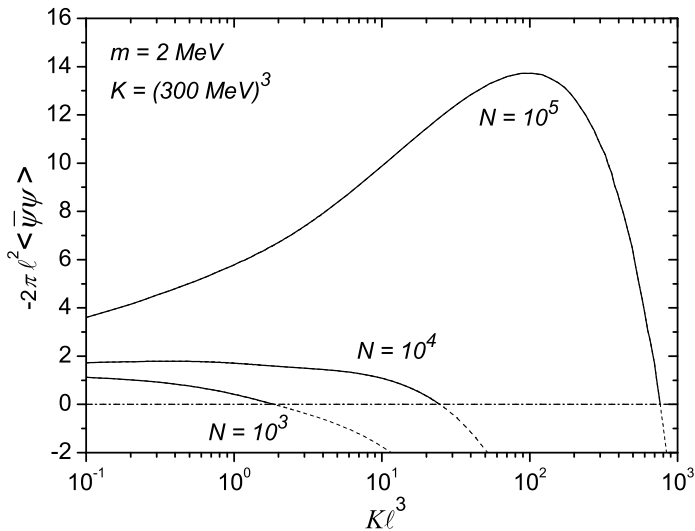}
\end{center}
\caption{Chiral condensate as a function of the dimensionless parameter $K
\ell^3$ for a current mass $m=2$~MeV, in the presence of a harmonic-oscillator
potential and a constant magnetic field in 2+1 dimensions.}
\label{condensatevschim2}
\end{figure}

In the presence of a nonvanishing current fermion mass, $m \neq 0$, the picture
is quite different. As an example, in Fig.~\ref{condensatevschim2} we present a
plot of the chiral condensate for $m=2$~MeV. Notice that the chiral condensate
diverges with the number $N$ of Landau levels and, consequently, a
renormalization scheme is needed for this case.

\section{Conclusion}
\label{sec:conclusion}

In this paper we have studied the problem of spontaneous chiral symmetry
breaking in the case of the simultaneous presence of a homogeneous magnetic
field and a fermionic quartic interaction, for both local and nonlocal kernels
in 2+1 and 3+1 dimensions. The operator formalism presented here is based on
the use of a sequence of Valatin-Bogoliubov canonical transformations, which
allowed us to construct explicitly the vacuum of the system and also to
calculate the chiral condensate. As illustrative examples, we have considered
the original NJL model and the harmonic-oscillator potential. In the case of
the NJL model we were able to recover the well-known mean-field results. For
the (2+1) harmonic oscillator, we have shown that in the presence of the
magnetic field the system (with nonlocal interaction) exhibits new properties,
namely, the existence of a critical magnetic field $B_{c}$ such that for
$B>B_c$ a chiral condensate is induced. It would interesting to extend this
analysis to the (3+1)-dimensional case. However, in the latter case the mass
gap equations turn out to be not only recurrent but also differential and,
therefore, their solution is more involved.

\section*{Acknowledgments}

We thank V.K. Dugaev for useful comments and suggestions. The work of
R.G.F. and G.M.M. has been supported by {\em Funda\c{c}\~{a}o para a
Ci\^{e}ncia e a Tecnologia} (FCT, Portugal) under the grants
SFRH/BPD/1549/2000 and SFRH/BD/984/2000, respectively.

\end{document}